\documentclass[%
reprint,
superscriptaddress,
aps,
amsmath,amssymb,
prl,
]{revtex4-1}

\usepackage{graphicx}
\usepackage{dcolumn}
\usepackage{bm}
\usepackage{braket}
\usepackage{framed}
\usepackage{setspace}

\begin{document}

\preprint{APS/123-QED}

\title{Recovering the full dimensionality of hyperentanglement in collinear photon pairs}

\author{Changjia Chen}
\email{changjia.chen@mail.utoronto.ca}
\author{Arash Riazi}
\author{Eric Y. Zhu}
\author{Li Qian}
\affiliation{Dept.of Electrical and Computer Engineering, University of Toronto, Toronto, M5S 3G4, Canada}

\date{\today}

\begin{abstract}
Exploiting hyperentanglement of photon pairs, that is, simultaneous entanglement in multiple degrees of freedom(DOFs), increases the dimensionality of Hilbert spaces for quantum information processing. However, generation of hyperentangled photon pairs collinearlly, while produces high brightness, results in a smaller Hilbert space due to the two photons being in the same spatial mode. In this letter, we point out that one can recover the full dimensionality of such hyperentanglement through a simple interference set up, similar to the time-reversed Hong-Ou-Mandel (TR-HOM) process. Different from the standard TR-HOM, we point out a critical phase condition has to be satisfied in order to recover the hyperentanglement. We theoretically analyze the realization of this approach and discuss the feasibility of generating truly hyperentangled photon pairs. Our proposed approach does not require post-selection and hence enables efficient hyper-entangled photon pairs generation for high-dimensional quantum applications.
\end{abstract}

\maketitle


Entangled photons play a critical role in many applications of quantum optics{horodecki2009quantum}. Photons that are simultaneously entangled in more than one degree of freedom(DOF), the so-called 'hyperentangled' states, have attracted much recent interest\cite{ ciampini2016path, HyperQDot2018Prilm,18QubitEntPanJW2018}. Hyperentanglement expands the  dimensionality of the Hilbert space of biphotons, enables complete Bell state analysis\cite{HyperCBSAKwiat2007,Superdensecoding2017}, increases the information capacity per pair photons\cite{barreiro2008beating, hu2018beating} and therefore lays the fundation of superdense-coding quantum communication\cite{steinlechner2017distribution, graham2015superdense,chapman2019time}. It has also become the key technology for certain tests of fundamental physics\cite{Zhang2019PRLbellselftest,Hyper2013GrahamPRLdynamics}. 

The generation of entangled photons is most conveniently done in a nonlinear medium. When the phase matching can be satisfied over a bandwidth much greater than the pump bandwidth, the photons are frequency entangled as a result of energy conservation. Entanglement in another DOF can be arranged through various means, such as type II phase matching for polarization-entanglement\cite{kwiat1995new}, delayed interferometer for time-bin entanglement\cite{reimer2019high}, or simultaneous phase matching for various orbital angular momentum modes \cite{zhang2016engineering}. Here, we consider hyper-entanglement in frequency and polarization DOFs, as frequency and polarization of photons are robust over large transmission distances in optical fibre. Entanglement in these two DOFs can also be generated relatively straightforwardly in fibre\cite{chen2017compensation} and nonlinear waveguides\cite{martin2010polarization}. While collinear photon pair generation in a fibre or a nonlinear waveguide is most efficient due to large nonlinear interaction length and no spatial filtering required (as opposed to non-collinear generation \cite{kwiat1997hyper}), we will show that that collinear entangled photon pairs have a reduced dimensionality in Hilbert space. 

By "hyperentanglement", we mean not only the photons are entangled in both DOFs, but also that they are \textit{accessible} individually, e.g. the hyperentangled photons are spatially separated.  In collinear parametric processes such as type-II spontaneous parametric down-conversion(SPDC), generally a state of entanglement in multiple DOFs can be generated, for example:
\begin{align}
\ket{\psi_{N}}=\frac{1}{\sqrt{2}}(\ket{H}\ket{V}+\ket{V}\ket{H})\otimes\sum_{n=1}^{N}\frac{1}{\sqrt{N}}\ket{\omega_{s,n}}\ket{\omega_{i,n}},\label{eqn.sub}
\end{align}
where $N$ is the total number of frequency-bin pairs. $\ket{H}$/$\ket{V}$ denotes the horizontally/vertically polarized polarized photon states and $\ket{\omega_{s/i,n}}$ denotes the state of signal/idler photon with center frequency $\omega_{s/i,n}$ which satisfies $\omega_{s,n}$+$\omega_{i,n}$=$const$\cite{xie2015harnessing,DiscreteColorEnt2009,lu2018quantum,Kaneda2019}. The generation of this state has been studied in many schemes, such as a two-period quasi-phase-matching nonlinear waveguide scheme\cite{Kaneda2019} and a cascaded photon pair source\cite{Riazi2019}. Notations of discretized frequency-bins are used in Eqn.(\ref{eqn.sub}) for the ease of discussion, but the results of this letter can also be applied to continuous frequency entanglement.  Note, due to collinear generation, the biphotons in the state $\ket{\psi_{N}}$ are in the same spatial mode. The accessible dimensionality of the biphoton states in collinear case is only 4$\times$$N$, in contrast to 8$\times$$N$ when the biphotons are in two different spatial modes(see supplementary material). To see the limitations of such state, consider the simplest case when $N=1$:
\begin{align}
\ket{\psi_{N=1}}=\frac{1}{\sqrt{2}}(\ket{H,\omega_s}\ket{V,\omega_i}+\ket{V,\omega_s}\ket{H,\omega_i}),\label{multiDOF}
\end{align}
If we separate the biphotons into two spatial modes by frequency, then frequency entanglement is destroyed, resulting in\cite{Kaneda2019,vergyris2017fully}: 
\begin{align}
\ket{\psi_{N=1,pol}}=\frac{1}{\sqrt{2}}(\ket{H}_1\ket{V}_2+\ket{V}_1\ket{H}_2)\otimes\ket{\omega_s}_1\ket{\omega_i}_2,\notag
\end{align}
If, on the other hand, we separate the biphotons by polarization, the polarization entanglement is destroyed\cite{Kaneda2019}:
\begin{align}
\ket{\psi_{N=1,frq}}=\frac{1}{\sqrt{2}}(\ket{\omega_s}_1\ket{\omega_i}_2+\ket{\omega_i}_1\ket{\omega_s}_2)\otimes\ket{H}_1\ket{V}_2\notag
\end{align}
where subscripts 1 or 2 refers to the spatial mode of the photonic state. A 50:50 beam splitter might also be used for probablistic separation of the biphotons, but in this case the PF hyperentanglement can only be obtained by post-selection in coincidence detection\cite{xie2015harnessing}. The state in Eqn.(\ref{multiDOF}) is not a direct product of two entangled states because both photons are in the same spatial mode. It is not genuine hyperentanglement, and for this reason we call it sub-hyperentanglement. In a non-collinear parametric process where the two photons are already in two spatial modes, there is only one proposal that we are aware of which produces genuine PF hyperentangled photon pairs\cite{kwiat1997hyper}. However, post-selection in spatial modes is required, resulting in inefficiency and difficulty in alignment.

In this letter we present a simple and novel scheme of PF hyperentanglement generation. By interfering two identical sub-hyperentangled photon pairs in Bell states from two coherently pumped nonlinear media on a beam splitter(BS) or polarizing beam splitter(PBS), one can generate PF hyperentanglement in two spatial modes of the form:
\begin{align}
\ket{\psi_{PF}}=&\frac{1}{\sqrt{2}}(\ket{H}_1\ket{V}_2+\ket{V}_1\ket{H}_2)\notag\\
\otimes\sum_{n=1}^{N}&\frac{1}{\sqrt{2N}}(\ket{\omega_{s,n}}_1\ket{\omega_{i,n}}_2+\ket{\omega_{i,n}}_1\ket{\omega_{s,n}}_2),\label{HyperEnt}
\end{align}
This scheme exploits the effect that two photon interference on a BS or PBS can deterministically separate biphotons in a pair\cite{TimereversePRAKumar2007,TimereversedHOM2018,marchildon2016deterministic}. By the interference of two two-mode squeezed states, one can realize a higher-dimensional entanglement on the biphotons that are already entangled and achieve PF hyperentanglement without the need for post-selection.
\begin{figure}[hbtp]
\centering
\includegraphics[width=7.5cm]{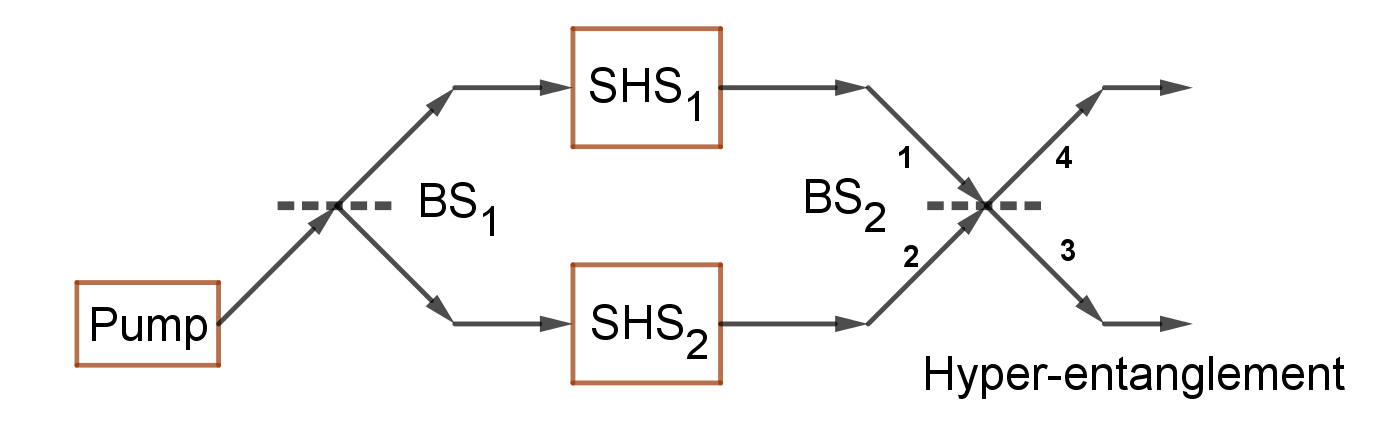}
\caption{Proposed scheme for polarization frequency hyper entanglement generation using a beam splitter: SHS, sub-hyperentanglement source; BS, beam splitter.}
\label{fig:setup}
\end{figure}

The proposed schematic setup is shown Fig.\ref{fig:setup}. A narrowband pump is split by a 50:50 BS and then sent to two identical nonlinear medium based sub-hyperentanglement sources(SHS). The sub-hyperentangled state $\ket{\psi_{N=1}}$ is prepared through nonlinear processes, where a two-mode squeezed state is generated. The states that arrive at the BS can be written as (without normalization)\cite{yang2008spontaneous}:
\begin{align}
\ket{\psi}_{k}\approx[1+\beta e^{i\theta_k}(\hat{a}^\dag_{H,s,k}\hat{a}^\dag_{V,i,k}+e^{i\phi_{RP}}\hat{a}^\dag_{V,s,k}\hat{a}^\dag_{H,i,k})]\ket{0},\label{hypo}
\end{align}
where $\ket{0}$ refers to vacuum, $\theta_k$ is the pump phase in path $k$=1 or 2, $a^{\dag}_{X,\alpha,k}$ is the creation operator of photon with frequency $\omega_\alpha$ ($\alpha=s$ or $i$) and polarization $X$=$H$ or $V$, and $\phi_{RP}\in[0,2\pi)$ is the relative phase between biphoton states $\hat{a}^\dag_{H,s,k}\hat{a}^\dag_{V,i,k}\ket{0}$ and $\hat{a}^\dag_{V,s,k}\hat{a}^\dag_{H,i,k}\ket{0}$. The coefficient $\beta\ll1$ is related to nonlinearity and phase-matching condition, and higher order terms of photon pair generation is neglected in weak pumping regime. For the ease of expression we only consider one frequency-bin pair here and the results can be extended to high-dimensional frequency-bins. We assume that the pump has a coherence length much longer than the optical path length difference of the two arms, such that the biphotons generated in the two SHS remain coherent and are able to interfere with each other in the rest of the setup\cite{CoherenceMandelPRL1991}. Assuming that the biphoton wavepackets from path 1 and 2 propagate through the same optical length and can sufficiently overlap on the BS in temporal/spatial domain, we can write the global input state on the BS as $\ket{\psi_{in}}=\ket{\psi}_1\ket{\psi}_2$. To avoid post-selection in detection, the two photons in the same input path has to be routed to two different output paths. It requires pump phase difference $\theta_2-\theta_1$ to be a multiple of $\pi$\cite{marchildon2016deterministic}. Without loss of generality we assume that $\theta_2-\theta_1$=0. Under the operation of the polarization preserving BS\cite{ou2007multi}: $\hat{a}_3=\frac{1}{\sqrt{2}}(\hat{a}_1+i\hat{a}_2)$, $\hat{a}_4=\frac{1}{\sqrt{2}}(i\hat{a}_1+\hat{a}_2)$, and by dropping the global phase, the second order terms which are proportional to $\beta^2$ and the vacuum, we obtain:
\begin{align}
\ket{\psi_{out}}&\approx\frac{1}{2}(\ket{H,\omega_s}_3\ket{V,\omega_i}_4+e^{i\phi_{RP}}\ket{H,\omega_i}_3\ket{V,\omega_s}_4\notag\\
&\quad+e^{i\phi_{RP}}\ket{V,\omega_s}_3\ket{H,\omega_i}_4+\ket{V,\omega_i}_3\ket{H,\omega_s}_4)\label{phasehyper}
\end{align}
where we use notations $\ket{X,\omega_\alpha}_k=a^{\dag}_{X,\alpha,k}\ket{0}$. As can be seen from Eqn.(\ref{phasehyper}), the biphotons from the same input are now in two different output paths because of interference. 

The deterministic separation of two photons by interference on BS can be interpreted as the time-reversed Hong-Ou-Mandel interference(TR-HOMI)\cite{TimereversePRAKumar2007}. In a regular TR-HOMI, any identical biphotons from two indistinguishable photon pair sources can be deterministically separated into two different outputs regardless of relative phase $\phi_{RP}$\cite{marchildon2016deterministic}. Yet previous reports do not result in hyperentanglement. To check the existence of hyperentanglement, we shall evaluate the entanglement in each DOF. A convenient measure of bipartite entanglement in density matrix $\rho$ is the concurrence $C(\rho)$, which is defined as $C(\rho)=max\{0,\lambda_1-\lambda_2-\lambda_3-\lambda_4\}$, $\lambda_i$ is the are the eigenvalues in decreasing order of the Hermitian matrix $R=\sqrt{\sqrt{\rho}\tilde{\rho}\sqrt{\rho}}$\cite{wootters1998entanglement}.  By calculating the partial density matries of $\rho_{out}=\ket{\psi_{out}}\bra{\psi_{out}}$ in polarization or frequency DOF through tracing over the other, we find the concurrences of biphotons in each DOF:
\begin{align}
C_{pol}[tr_{\omega}(\rho_{out})] = C_{\omega}[tr_{pol}(\rho_{out})]=|\cos\phi_{RP}|
\end{align}
Only when $\phi_{RP}$ = 0 or $\pi$ can the concurrence in both DOFs become 1 and maximal entanglement be obtained. For instance, when $\phi_{RP}$ = 0, we have:
\begin{align}
\ket{\psi_{out}}=\frac{1}{2}&(\ket{H}_3\ket{V}_4+\ket{V}_3\ket{H}_4)\notag\\
&\otimes(\ket{\omega_s}_3\ket{\omega_i}_4+\ket{\omega_i}_3\ket{\omega_s}_4),\label{outputbs}
\end{align}
The output state as shown Eqn.(\ref{outputbs}) can then be truly PF hyperentangled, which is identical to Eqn.(\ref{HyperEnt}) with $N=1$.

Besides using a BS, a PBS can also be used to ``hyperentangle'' entangled photons by mapping polarizations to spatial modes. Assuming that a PBS is frequency-independent, it can deterministically separate biphotons by reflecting H- and transmitting V-polarized photons regardless of the pump phase difference $\theta_1-\theta_2$. The input $\ket{\psi_{in}}=\ket{\psi}_1\ket{\psi}_2$ in path 1 and 2 of PBS will result in a similar output state as Eqn.(\ref{phasehyper}). With the same phase requirement $\phi_{RP}$ = 0 or $\pi$ satisfied, we are able to obtain polarization and frequency entanglement simultaneously in two spatial modes with a PBS.

The nontrivial difference between our proposal and previous interferometers for entangled photon generation\cite{TimereversePRAKumar2007, TimereversedHOM2018,marchildon2016deterministic,yoshizawa2004generation,yoshizawa2003generation} is that our proposed setup uses the photon pairs in sub-hyperentangled Bell states as input. As mentioned, the sub-hyperentangled state Eqn.(\ref{multiDOF}) is only entangled in \textit{either} polarization \textit{or} frequency DOF. The operation of BS or PBS ``entangles'' the unentangled states in an additional DOF of the biphotons, by superposing two two-mode squeezed states and routing them into different spatial modes. As is shown in Eqn.(\ref{phasehyper}), the existence of $\phi_{RP}$ leads to correlation between polarization and frequency DOF, such that the entanglement in each subspace might be lost. We point out that by keeping $\phi_{RP}=0$ or $\pi$, we can generate the genuine hyperentanglement of two DOFs in two spatial modes without affecting the existing entanglement and realize more accessible dimensionalities . 

\begin{figure}[ht]
\centering
\includegraphics[width=8.5cm]{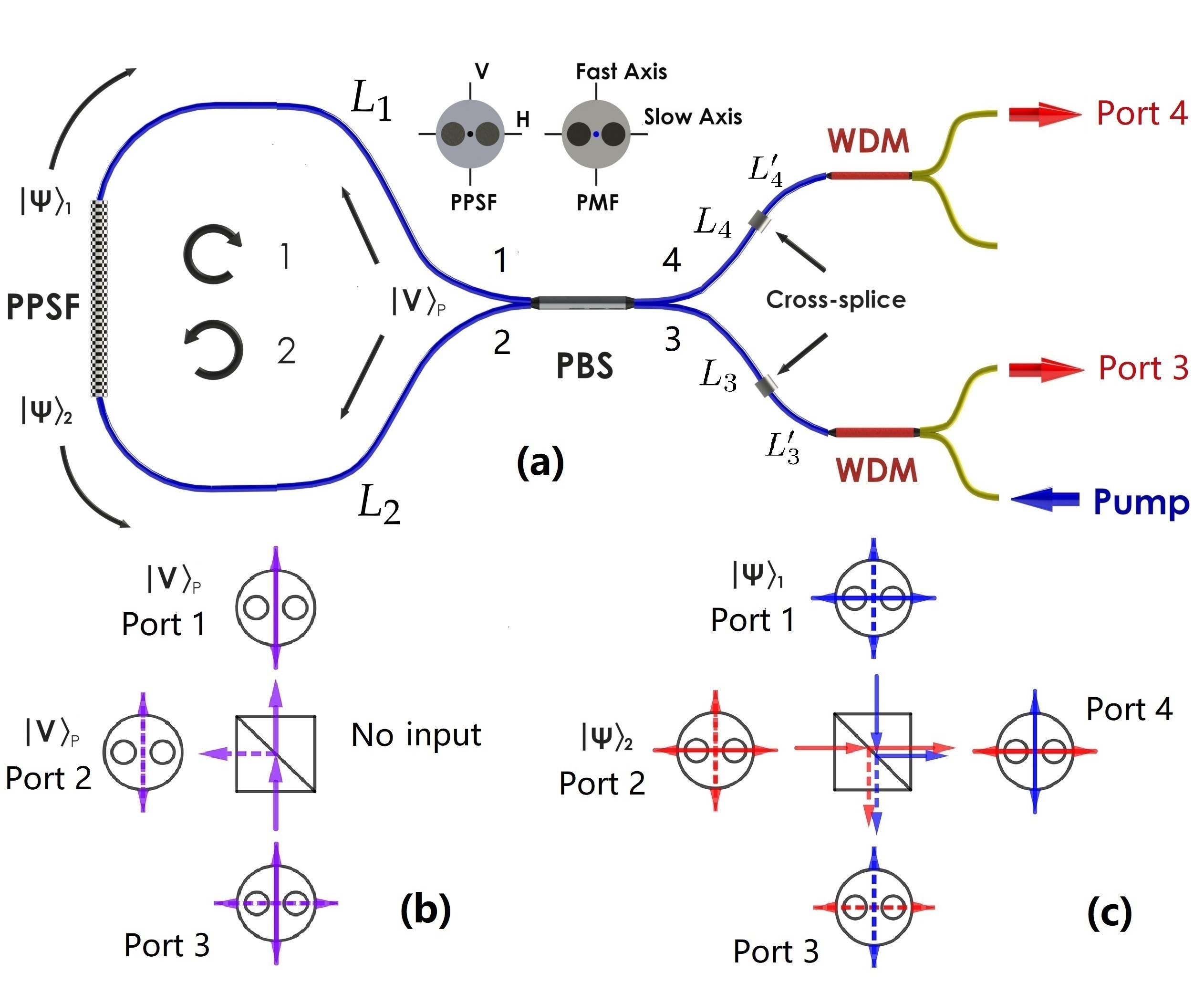}
\caption{(a)Proposed polarization frequency hyperentangled biphoton source using periodically poled silica fibre(PPSF). WDM, wavelength division multiplexer for 780nm/1550nm. PBS, polarizing beam splitter. Blue lines are polarization maintaining fibre. Yellow lines are single mode fibre. The inset shows the correspondence of polarization modes bewteen PPSF and PMF. (b) Polarization transformation of pump in fibre-coupled inline PBS; (c) Polarization transformation of down-converted biphotons in fibre-coupled inline PBS;}
\label{fig:PPSFsetup}
\end{figure}

A Mach-Zehnder interferometer as proposed in Fig.\ref{fig:setup} requires precise active phase stabilization. An alternative realization of PF hyperentanglement that avoids additional phase stabilization is using a Sagnac loop phase-stable interferometer scheme in combination with a PBS/BS that can function at all of the pump, signal and idler frequencies\cite{kim2006phase}. As an example, here we present a design of a periodically poled silica fibre(PPSF) based Sagnac loop setup for PF hyperentanglement photon pair generation. 
 
As shown in Fig.\ref{fig:PPSFsetup}a, A pump laser is sent through a 780nm/1550nm wavelength division multiplexer(WDM) and then enters the Sagnac loop through a fibre-coupled PBS. The pump light that is directed to clockwise direction (path 1) and counter-clockwise direction (path 2) in the Sagnac loop will both propagate at the fast axis of polarization maintaining fibre(PMF) as shown in Fig.\ref{fig:PPSFsetup}b. A PPSF is used for entanglement generation via type-II SPDC, in which a vertically polarized pump photon is down-converted to a pair of H- and V- polarized photons\cite{zhu2012direct}. The fast axis of PMF is aligned to PPSF's V direction(inset of Fig.\ref{fig:PPSFsetup}a)\cite{chen2018turn}. Owing to the PPSF's low group birefringence nature, it can offer high quality direct generation of entangled photons states Eqn.(\ref{hypo}) with $\phi_{RP}=0$ when PPSF is properly pumped(for example, at $>$0.6nm of pump detuning wavelength as shown in Fig.3 of \cite{chen2017compensation}) or spectrally filtered. The generated pairs then go through PMF $L_1$ or $L_2$ at each end of PPSF as denoted in Fig.\ref{fig:PPSFsetup}a and are mixed on the PBS. We assume the polarization transformation in a off-the-shelf PMF-coupled PBS is can be described as follow(Fig.\ref{fig:PPSFsetup}c): $
\ket{H}_1$$\rightarrow$$\ket{V}_4$, $\ket{V}_1$$\rightarrow$$\ket{V}_3$, $\ket{H}_2$$\rightarrow$$\ket{H}_4$, $\ket{V}_2$$\rightarrow$$\ket{H}_3$.
At the output path 3 and 4, there are additional PMFs  $L_3$ and $L_4$ connected to the PBS. Because of birefringence in $L_1$$\sim$$L_4$, additional cross-spliced PMFs after $L_3$ and $L_4$ are needed for compensation, which are labelled by $L_3'$ and $L_4'$ respectively.  We may then write the biphoton output states at port 3 and 4 as(see supplementary material Eqn.(S2)):
\begin{widetext}
\begin{align}
\begin{split}
\ket{\psi_{PBS, out}}=\frac{1}{2}\bigg\{&\ket{H,\omega_s}_3\ket{H,\omega_i}_4e^{i[k_pL_2+k_V(\omega_s)L_1+k_H(\omega_i)L_1+k_V(\omega_s)L_3+k_H(\omega_s)L_3'+k_V(\omega_i)L_4+k_H(\omega_i)L_4']}\\
+&\ket{H,\omega_i}_3\ket{H,\omega_s}_4e^{i[k_pL_2+k_H(\omega_s)L_1+k_V(\omega_i)L_1+k_V(\omega_i)L_3+k_H(\omega_i)L_3'+k_V(\omega_s)L_4+k_H(\omega_s)L_4']}\\
+&\ket{V,\omega_s}_3\ket{V,\omega_i}_4e^{i[k_pL_1+k_V(\omega_s)L_2+k_H(\omega_i)L_2+k_H(\omega_s)L_3+k_V(\omega_s)L_3'+k_H(\omega_i)L_4+k_V(\omega_i)L_4']}\\
+&\ket{V,\omega_i}_3\ket{V,\omega_s}_4e^{i[k_pL_1+k_H(\omega_s)L_2+k_V(\omega_i)L_2+k_H(\omega_i)L_3+k_V(\omega_i)L_3'+k_H(\omega_s)L_4+k_V(\omega_s)L_4']}\bigg\}
\end{split}
\end{align}
\end{widetext}
where $k_p$ is the wavenumber of the pump light, $k_{H/V}(\omega)$ is the PMF's wavenumber in H/V polarization at frequency $\omega$. To see the simultaneous entanglement in both DOFs, we calculate the concurrence of $\rho_{PBS, out}=\ket{\psi_{PBS,out}}\bra{\psi_{PBS,out}}$ in each DOFs:
\begin{widetext}
\begin{align}
C_{pol}[tr_{\omega}(\rho_{PBS,out})] = C_{\omega}[tr_{pol}(\rho_{PBS,out})]=\left|\cos\left\{\frac{1}{2}[k_H(\omega_s)-k_V(\omega_s)+k_V(\omega_i)-k_H(\omega_i)]\Delta L\right\}\right|\label{conPBS}
\end{align} 
\end{widetext}
where $\Delta L=|L_1-L_2+L_3-L_4-L_3'+L_4'|$ shall be zero in order to achieve hyperentanglement. Eqn.(\ref{conPBS}) shows that a non-zero $\Delta L$ might lead to polarization or frequency decoherence. In addition, though the usage of PMF prevents polarization misalignment, its birefringence will lead to temporal decoherence because of the temporal walk-off between H- and V- polarized photons(see supplementary material Eqn.(S7)-(S10)). In order to compensate for the temporal decoherence, additional condition on the length of PMF needs to be satisfied(supplementary material Eqn.(S14)):
\begin{widetext}
\begin{align}
(L_1+L_2)(\frac{1}{v_{gV}}-\frac{1}{v_{gH}})+(L_3+L_3'-L_4-L_4')(\frac{1}{v_{gV}}+\frac{1}{v_{gH}})=0
\end{align}
\end{widetext}
where $v_{gV}$ and $v_{gH}$ are the group velocities in the fast and slow axes of PMF. The length error of PMFs has to be sufficiently small. Experimentally, the length error can be controlled to $<$1cm\cite{vergyris2017fully}, which are sufficient for maintaining coherence(see supplementary material).  The hyperentanglement can thus maintain high entanglement in each DOF at the output. In this all-fibre scheme, the requirements of identical SHS and phase stabilization are relieved. It is compact, robust, and therefore an enabler for a broader range of quantum applications.

In conclusion, we proposed a simple and novel scheme of increasing the accessible dimensionality of hyperentanglement by interfering two identical sub-hyperentangled biphoton Bell states. The interference can deterministically seperate the photons in a pair and achieve hyperentanglement in polarization and frequency DOF without post-selection if the phase requirement is satisfied. We show that such PF hyperentanglement will provide more accessible dimensionalities and we propose a feasible experiment design. This novel approach reduces the need of filtering or post-selection in detection and enables efficient PF hyperentanglement generation in waveguide using the widely used nonlinear medium platform. It can be applied to high dimensional discrete frequency-bin or continuous frequency entangled states. This approach may provide a route towards the generation of higher-dimensional hyperentanglement and its applications in quantum information processing.

\bibliography{sample}

\begin{widetext}
\section*{Supplementary}
\subsection*{Counting the accessible dimensionality of polarization-frequency hyperentanglement}
Due to energy conservation, the sum of the biphoton frequencies has to be constant, i.e. $\omega_{s,n}$+$\omega_{i,n}$=const, and all the other biphoton frequency states such as $\ket{\omega_{s,n}}\ket{\omega_{s,n}}$ and $\ket{\omega_{i,n}}\ket{\omega_{i,n}}$ cannot be accessible. In co-linear case, the dimensionality of polarization$\otimes$frequency will be 4(polarization)$\times$N(frequency), where in frequency domain one could only access to basis \{$\ket{\omega_{s,n}}\otimes\ket{\omega_{i,n}}$\}. In two spatial modes case, as one could access to basis \{$\frac{1}{\sqrt{2}}(\ket{\omega_{s,n}}_1\ket{\omega_{i,n}}_2\pm\ket{\omega_{i,n}}_1\ket{\omega_{s,n}}_2)$\}, the dimensionality becomes 4(polarization)$\times$2N(frequency)=8$\times$N.

\subsection*{Calculation of the output state and the corresponding concurrence in Sagnac loop polarization-frequency hyperentanglement source}
In the following context we will show the calculation of the output state of the Sagnac loop hyperenentanglement and the estimation of the entanglement quality in it. All of the notations in the following supplementary are consistent with the main text. First of all, we may write the biphoton state at the output ends of PPSF inside the Sagnac loop:
\begin{align}
\ket{\psi_1}=\frac{1}{\sqrt{2}}e^{ik_pL_2}\iint\mathrm{d}\omega_a\mathrm{d}\omega_bf(\omega_a,\omega_b)[a^{\dag}_{H,1}(\omega_a)a^{\dag}_{V,1}(\omega_b)+a^{\dag}_{V,1}(\omega_a)a^{\dag}_{H,1}(\omega_b)]\ket{0},\notag\\
\ket{\psi_2}=\frac{1}{\sqrt{2}}e^{ik_pL_1}\iint\mathrm{d}\omega_a\mathrm{d}\omega_bf(\omega_a,\omega_b)[a^{\dag}_{H,2}(\omega_a)a^{\dag}_{V,2}(\omega_b)+a^{\dag}_{V,2}(\omega_a)a^{\dag}_{H,2}(\omega_b)]\ket{0},\notag
\end{align}
where $\omega_a$ and $\omega_b$ are dummy variables, $f(\omega_a,\omega_b)$ is the joint spectral amplitude function and $\iint\mathrm{d}\omega_a\mathrm{d}\omega_b|f(\omega_a,\omega_b)|^2=1$. We assume that pump light is narrow linewidth, such that $\omega_a+\omega_b\equiv\omega_p$. We may then have rewrite the states into:
\begin{align}
\begin{split}
\ket{\psi_1}=\frac{1}{\sqrt{2}}e^{ik_pL_2}\int\mathrm{d}\omega g(\omega)[a^{\dag}_{H,1}(\omega)a^{\dag}_{V,1}(\omega_p-\omega)+a^{\dag}_{V,1}(\omega)a^{\dag}_{H,1}(\omega_p-\omega)]\ket{0},\notag\\
\ket{\psi_2}=\frac{1}{\sqrt{2}}e^{ik_pL_1}\int\mathrm{d}\omega g(\omega)[a^{\dag}_{H,2}(\omega)a^{\dag}_{V,2}(\omega_p-\omega)+a^{\dag}_{V,2}(\omega)a^{\dag}_{H,2}(\omega_p-\omega)]\ket{0},
\end{split}\tag{S1}
\end{align}
where $g(\omega)\propto f(\omega,\omega_p-\omega)$ and $\int\mathrm{d}\omega|g(\omega)|^2=1$. 
\begin{itemize}
\item{Narrowband frequency-bin approximation}
\end{itemize}

Considering a pair of frequency-bins that have center frequencies $\omega_s$ and $\omega_i$ ($\omega_s>\omega_i$, $\omega_s+\omega_i=\omega_p$), in the ideal case we assume that the bandwidth of frequency-bins are infinitely narrow, i.e., $g(\omega)\approx\delta(\omega-\omega_s)$. We can use the ket notation $\ket{X,\omega_\alpha}_k=\ket{X}_k\otimes\ket{\omega_\alpha}_k=a^{\dag}_{X,k}(\omega_\alpha)\ket{0}$ ($X$=$H$ or $V$, $\alpha=$$s$ or $i$) and rearrange Eqn.(S1):
\begin{align}
\ket{\psi_1}=\frac{1}{\sqrt{2}}e^{ik_pL_2}(\ket{H,\omega_s}_1\ket{V,\omega_i}_1+\ket{V,\omega_s}_1\ket{H,\omega_i}_1),\notag\\
\ket{\psi_2}=\frac{1}{\sqrt{2}}e^{ik_pL_1}(\ket{H,\omega_s}_2\ket{V,\omega_i}_2+\ket{V,\omega_s}_2\ket{H,\omega_i}_2).\notag
\end{align}
After being generated from PPSF, the states $\ket{\psi_1}$ and $\ket{\psi_2}$ propagate through PMF $L_1$ and $L_2$ respectively, and are mixed in PBS though transformation: $
\ket{H}_1\rightarrow\ket{V}_4,\ \ket{V_1}\rightarrow\ket{V}_3,\ \ket{H}_2\rightarrow\ket{H}_4, \ \ket{V}_2\rightarrow\ket{H}_3$. We can find the state at the output of PBS:
\begin{align}
\ket{\psi_{PBS}}=\frac{1}{2}\bigg\{&\ket{H,\omega_s}_3\ket{H,\omega_i}_4e^{i[k_pL_1+k_V(\omega_s)L_2+k_H(\omega_i)L_2]}+\ket{H,\omega_i}_3\ket{H,\omega_s}_4e^{i[k_pL_1+k_H(\omega_s)L_2+k_V(\omega_i)L_2]}\notag\\
+&\ket{V,\omega_s}_3\ket{V,\omega_i}_4e^{i[k_pL_2+k_V(\omega_s)L_1+k_H(\omega_i)L_1]}
+\ket{V,\omega_i}_3\ket{V,\omega_s}_4e^{i[k_pL_2+k_H(\omega_s)L_1+k_V(\omega_i)L_1]}\bigg\}\notag
\end{align}
The output ports 3 and 4 of PBS are also assumed to be coupled with PMFs of length $L_3$ and $L_4$ respectively. In order to compensate for birefringence, additional PMFs $L_3'$ and $L_4'$ are crossed spliced to PMFs $L_3$ and $L_4$, which map $\ket{H}_{k}\xrightarrow{\mathrm{cross\ splice}}\ket{V}_{k}$, $k$=3 or 4. The output state that reaches at the output of $L_3'$ and $L_4'$ will be:
\begin{align}
\ket{\psi_{PBS}'}=\frac{1}{2}\bigg\{&\ket{H,\omega_s}_3\ket{H,\omega_i}_4e^{i[k_pL_2+k_V(\omega_s)L_1+k_H(\omega_i)L_1+k_V(\omega_s)L_3+k_H(\omega_s)L_3'+k_V(\omega_i)L_4+k_H(\omega_i)L_4']}\notag\\
+&\ket{H,\omega_i}_3\ket{H,\omega_s}_4e^{i[k_pL_2+k_H(\omega_s)L_1+k_V(\omega_i)L_1+k_V(\omega_i)L_3+k_H(\omega_i)L_3'+k_V(\omega_s)L_4+k_H(\omega_s)L_4']}\notag\\
+&\ket{V,\omega_s}_3\ket{V,\omega_i}_4e^{i[k_pL_1+k_V(\omega_s)L_2+k_H(\omega_i)L_2+k_H(\omega_s)L_3+k_V(\omega_s)L_3'+k_H(\omega_i)L_4+k_V(\omega_i)L_4']}\tag{S2}\\
+&\ket{V,\omega_i}_3\ket{V,\omega_s}_4e^{i[k_pL_1+k_H(\omega_s)L_2+k_V(\omega_i)L_2+k_H(\omega_i)L_3+k_V(\omega_i)L_3'+k_H(\omega_s)L_4+k_V(\omega_s)L_4']}\bigg\}\notag
\end{align}

The above equation is shown in main text as Eqn.(8). As the frequency-bins are narrow linewidth, we can trace over the computational basis $\{\ket{\omega_s}_3\ket{\omega_s}_4,\ket{\omega_s}_3\ket{\omega_i}_4,\ket{\omega_i}_3\ket{\omega_s}_4,\ket{\omega_i}_3\ket{\omega_i}_4\}$ then write the partial density matrix in polarization as:
\begin{align}
\rho_{pol}=tr_\omega(\ket{\psi_{PBS}'}\bra{\psi_{PBS}'})=\frac12\left(\begin{matrix}
1 & 0 & 0 & \frac{1}{2}(e^{i\phi_A}+e^{i\phi_B})\\
0 & 0 & 0 & 0\\
0 & 0 & 0 & 0\\
\frac{1}{2}(e^{-i\phi_A}+e^{-i\phi_B}) &0 &0 &1
\end{matrix}\right)\notag
\end{align}
where the phase $\phi_A$ and $\phi_B$ is 
\begin{align}
\phi_A = [k_p-k_V(\omega_s)-k_H(\omega_i)](L_2-L_1)+[k_V(\omega_s)-k_H(\omega_s)](L_3-L_3')+[k_V(\omega_i)-k_H(\omega_i)](L_4-L_4'),\notag\\
\phi_B = [k_p-k_H(\omega_s)-k_V(\omega_i)](L_2-L_1)+[k_V(\omega_i)-k_H(\omega_i)](L_3-L_3')+[k_V(\omega_s)-k_H(\omega_s)](L_4-L_4').\notag
\end{align}
And we can also calculate the partial density matrix in frequency by tracing over the polarization DOF:
\begin{align}
\rho_{\omega}=tr_{pol}(\ket{\psi_{PBS}'}\bra{\psi_{PBS}'})=\frac12\left(\begin{matrix}
0 & 0 & 0 & 0\\
0 & 1 & \frac{1}{2}(e^{i\phi_C}+e^{i\phi_D}) & 0\\
0 & \frac{1}{2}(e^{-i\phi_C}+e^{-i\phi_D}) & 1 & 0\\
0 &0 &0 &0
\end{matrix}\right)\notag
\end{align}
where the phase $\phi_C$ and $\phi_D$ is 
\begin{align}
\phi_C =[k_V(\omega_s)+k_H(\omega_i)-k_H(\omega_s)-k_V(\omega_i)]L_2+[k_H(\omega_s)-k_H(\omega_i)](L_3-L_4)+[k_V(\omega_s)-k_V(\omega_i)](L_3'-L_4'),\notag\\
\phi_D =[k_V(\omega_s)+k_H(\omega_i)-k_H(\omega_s)-k_V(\omega_i)]L_1+[k_V(\omega_s)-k_V(\omega_i)](L_3-L_4)+[k_H(\omega_s)-k_H(\omega_i)](L_3'-L_4').\notag
\end{align}
The corresponding concurrences are: 
\begin{align}
C(\rho_{pol})&=\left|\cos[\frac12(\phi_A-\phi_B)]\right|,\quad C(\rho_{\omega})=\left|\cos[\frac12(\phi_C-\phi_D)]\right|, \tag{S3}\label{concurrencerho}
\end{align}
where $|\phi_A-\phi_B|=|\phi_C-\phi_D|=|[k_H(\omega_s)-k_V(\omega_s)+k_V(\omega_i)-k_H(\omega_i)]\Delta L|$, and
$\Delta L=L_1-L_2+L_3-L_4-L_3'+L_4'$. The Eqn.(\ref{concurrencerho}) is shown as Eqn.(9) in the main text. As can be shown above, when $\Delta L=0$, we can obtain maximally entanglement in both polarization and frequency DOFs and achieve PF hyperentanglement.

In practice, to estimate the phase inside the cosine argument of Eqn.(S3), we can do Taylor expansion on $k_X(\omega)$ ($X=$H or V) around the center wavelength $\omega_{c}=\frac{1}{2}(\omega_s+\omega_i)$:
\begin{align}
k_X(\omega)=k_{0X}+\frac{1}{v_{gX}}(\omega-\omega_{c})+\frac{1}{2}k_{2X}(\omega-\omega_{c})^2+...\tag{S4}\label{Taylor}
\end{align}
where $k_{0X}=k_X(\omega_{c})$, $\frac{1}{v_{gX}}=\left.\frac{\mathrm{d}k_X}{\mathrm{d}\omega}\right|_{\omega=\omega_{c}}$ is the inverse of group velocity in X polarization, and $k_{2X}=\left.\frac{\mathrm{d}^2k_X}{\mathrm{d}\omega^2}\right|_{\omega=\omega_{c}}$ is group velocity dispersion at $\omega_{c}$. By ignoring any terms that are higher than the second order, we find the phase argument in Eqn.(S3) is 
\begin{align}
|[k_H(\omega_s)-k_V(\omega_s)+k_V(\omega_i)-k_H(\omega_i)]\Delta L|\approx\left|(\frac{1}{v_{gV}}-\frac{1}{v_{gH}})(\omega_s-\omega_i)\Delta L\right|. \notag
\end{align}
In the above equation we make an approximation that $\frac12|(\beta_{2H}-\beta_{2V})(\omega-\omega_c)^2|\ll|(\frac{1}{v_{gV}}-\frac{1}{v_{gH}})(\omega-\omega_c)|$\cite{okamoto1987polarization}. Note that the group velocity mismatch $M$ bewteen two orthorgonal principal polarizations of PMF can be approximated:
\begin{align}
M=\frac{1}{v_{gV}}-\frac{1}{v_{gH}}=\left(\frac{\Delta n}{c}+\frac{\omega_{c}}{c}\frac{\mathrm{d}\Delta n}{\mathrm{d}\omega}\right)\approx\frac{\lambda}{cL_B}\tag{S5}\label{GVM}
\end{align}
where $\Delta n$ is the modal birefringence, $L_B$ is the beat length of PMF, and $\frac{\mathrm{d}\Delta n}{\mathrm{d}\omega}\approx0$. The beat length of standard PMF can be found online, for example, $\sim$4mm see Ref.\cite{CorningPM}. As a rough estimation, we assume a frequency spacing $\omega_s-\omega_i=2\pi\times2$THz and a center wavelength of $\lambda$=1550nm. To obtain the concurrence as shown in Eqn.(S3) of higher than 0.999 (0.99), we will need $\Delta L<0.55$cm (1.74cm), which is experimentally feasible. 

\begin{itemize}
\item{Finite linewidth frequency-bins}
\end{itemize}
\subsubsection{Polarization decoherence}
Practically the spectrum of frequency-bins are not kronecker $\delta$ functions and it shall have a finite bandwidth. We keep the integral in Eqn.(S1) and write the output state that arrives at the outputs of PMF $L_3'$ and $L_4'$ as:
\begin{align}
\begin{split}\ket{\psi'_{PBS}}=\frac{1}{2}\int_{\omega>\omega_c}\mathrm{d}\omega g(\omega)[&a^{\dag}_{H,3}(\omega)a^{\dag}_{H,4}(\omega_p-\omega)e^{i[k_pL_2+k_V(\omega)L_1+k_H(\omega_p-\omega)L_1+k_V(\omega)L_3+k_H(\omega)L_3'+k_V(\omega_p-\omega)L_4+k_H(\omega_p-\omega)L_4']}\notag\\
+&a^{\dag}_{H,3}(\omega_p-\omega)a^{\dag}_{H,4}(\omega)e^{i[k_pL_2+k_H(\omega)L_1+k_V(\omega_p-\omega)L_1+k_V(\omega_p-\omega)L_3+k_H(\omega_p-\omega)L_3'+k_V(\omega)L_4+k_H(\omega)L_4']}\\
+&a^{\dag}_{V,3}(\omega)a^{\dag}_{V,4}(\omega_p-\omega)e^{i[k_pL_1+k_V(\omega)L_2+k_H(\omega_p-\omega)L_2+k_H(\omega)L_3+k_V(\omega)L_3'+k_H(\omega_p-\omega)L_4+k_V(\omega_p-\omega)L_4']}\notag\\
+&a^{\dag}_{V,3}(\omega_p-\omega)a^{\dag}_{V,4}(\omega)e^{i[k_pL_1+k_H(\omega)L_2+k_V(\omega_p-\omega)L_2+k_H(\omega_p-\omega)L_3+k_V(\omega_p-\omega)L_3'+k_H(\omega)L_4+k_V(\omega)L_4']}]\ket{0}\notag
\end{split}\tag{S6}\label{CVState}
\end{align}
The partial density matrix in polarization can be calculated by:
\begin{align}
\rho_{pol}=\iint_{\omega',\omega''}\bra{0}\mathrm{d}\omega'\mathrm{d}\omega''a_3(\omega')a_4(\omega'')\ket{\psi'_{PBS}}\bra{\psi'_{PBS}}a_3^{\dag}(\omega')a_4^{\dag}(\omega'')\ket{0}
=\frac{1}{2}\left(\begin{matrix}
1 & 0 & 0 & \alpha\\
0 & 0 & 0 & 0\\
0 & 0 & 0 & 0\\
\alpha* &0 &0 & 1
\end{matrix}\right)\notag
\end{align}
where we use notation $\ket{\omega}_k=a^{\dag}_k(\omega)\ket{0}$ and the matrix element 
\begin{align}
\alpha=\frac12\int_{\omega>\omega_c}\mathrm{d}\omega g(\omega)g^*(\omega)\big\{&e^{i[k_p-k_V(\omega)-k_H(\omega_p-\omega)](L_2-L_1)+[k_V(\omega)-k_H(\omega)](L_3-L_3')+[k_V(\omega_p-\omega)-k_H(\omega_p-\omega)](L_4-L_4')}\notag\\
+&e^{i[k_p-k_H(\omega)-k_V(\omega_p-\omega)](L_2-L_1)+[k_V(\omega_p-\omega)-k_H(\omega_p-\omega)](L_3-L_3')+[k_V(\omega)-k_H(\omega)](L_4-L_4')}\big\}\notag,
\end{align}
and $\alpha^*$ is its complex conjugate. When $\Delta L=0$, the maximal polarization entanglement is obtained, which agrees with the narrow band assumption.

\begin{figure}[tbp]
\centering
\includegraphics[width=12cm]{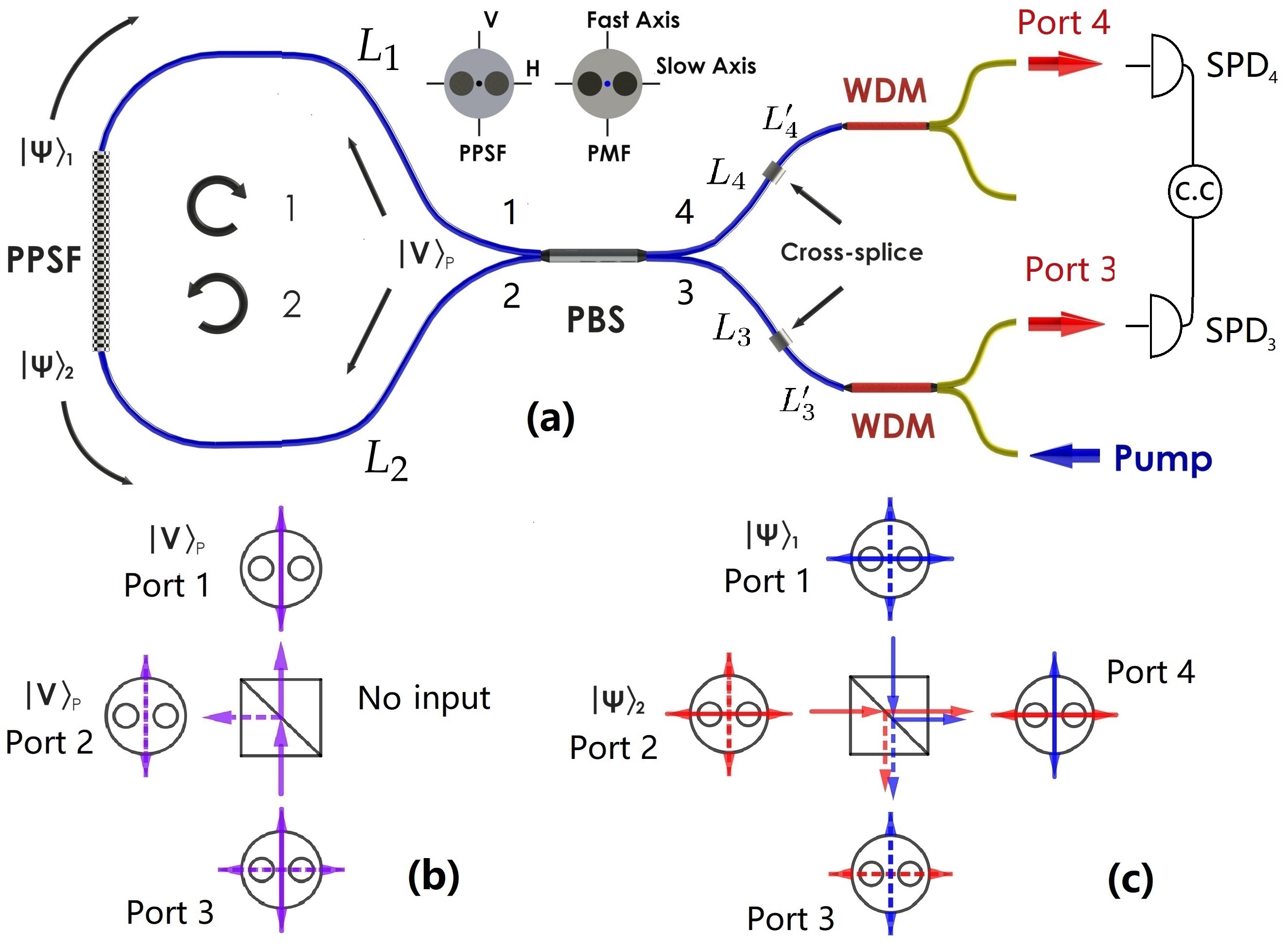}
\caption{(a)Proposed polarization frequency hyperentangled biphoton source using periodically poled silica fiber(PPSF). WDM, wavelength division multiplexer for 780nm/1550nm. PBS, polarizing beam splitter, PA, polarization analyzer. SPD, single photon detectors. CC, coincidence counter. Blue lines are polarization maintaining fiber. Yellow lines are single mode fiber. The inset shows the correspondence of polarization modes bewteen PPSF and PMF. (b) Polarization transformation of pump in fiber-coupled inline PBS; (c) Polarization transformation of down-converted biphotons in fiber-coupled inline PBS;}
\label{fig:PPSFsetup}
\end{figure}

\subsubsection{Frequency and temporal decoherence}
Since we assume a finite linewidth in frequency-bins, the partial density matrix in frequency DOF is no longer describing two qubits states. Thus the ``concurrence'' which provides the measure of entanglement of two qubits will not be applicable in this case. Note that the pump is assumed to be narrow bandwidth and energy conservation always holds, and Eqn.(S2)-(S3) hold for every single frequency-pair $\{\delta(\omega-\omega_s),\delta(\omega-\omega_i)\}$. When $\Delta L=0$, the maximal PF hyperentanglement can still be achieved in every single frequency-pair. The only difference between narrow bandwidth frequency-bins and finite bandwidth frequency-bins is that there will be frequency dependent components in the phase arguments. The frequency dependent components will be revealed in temporal domain upon detection, or in other words, lead to temporal decoherence. In the following discussion we will show the evaluation in temporal domain.

To evaluate the temporal behavior of the PF hyperentangled photons, we shall analyze the biphoton wavefunction $\Psi(t_3,t_4)$  following the methods in Ref.\cite{shih1994two}:
\begin{align}
\Psi(t_3,t_4)=\bra{0}E_3^{(+)}(t_3)E_4^{(+)}(t_4)\ket{\psi'_{PBS}}\tag{S7}\label{tfunction}
\end{align}
where $E_3^{(+)}(t)$ and $E_4^{(+)}(t)$ are the fields at the detector 3 and 4 which are placed at the output port 3 and 4 after WDM, as shown in Fig.1. The time variables $t_3$ and $t_4$ are the detection time of correponding detectors. The fields $E_3^{(+)}(t)$ and $E_4^{(+)}(t)$ can be written as:
\begin{align}
\begin{split}
E_3^{(+)}(t)=E\int\mathrm{d}\omega h_3(\omega)\exp[-i\omega(t-\tau_3)]\sum_X\hat{e_3}\centerdot\hat{e_X}a_{X,3}(\omega),\\
E_4^{(+)}(t)=E\int\mathrm{d}\omega h_4(\omega)\exp[-i\omega(t-\tau_4)]\sum_X\hat{e_4}\centerdot\hat{e_X}a_{X,4}(\omega),
\end{split}
\tag{S8}\label{detectorfield}
\end{align}
where $X$= $H$ or $V$ is the polarization of photons, $\hat{e_i}$ is in the direction of the $i$th linear polarization analyzer axis, $h_i(\omega)$ is the spectral transmission function of the filter in front of $i$th detector, and $\tau_i$ is the travelling time of light from the output of PMF $L_i'$ to detector $i$=3 or 4, and $E$ is a constant. 

For ease of calculation, we make following assumptions and approximations:

1, As the delay to detectors are adjustable, we can assume $\tau_3=\tau_4=0$. 

2, We assume a CW-pump is used such that the pump spectrum is taken to be $\delta(\omega-\omega_p)$ and the overall fields of down-converted photons are infinitely long in temporal domain. With this assumption the down-converted photons from different input ports 1 and 2 can sufficiently overlap in time on the PBS, and the pump phase terms such as $k_pL_1$ and $k_pL_2$ in Eqn.(\ref{CVState}) will not lead to additional temporal decoherence. This can be easily achieve by using a pump with coherence length much longer than $|L_1-L_2|$.

3, For the ease of calculation we can assume that the biphoton spectral amplitude function $g(\omega)$ is taken to be a pair of flat-top frequency bins with center wavelengths $\omega_s$ and $\omega_i$, and with bandwidth of $\Delta\omega$:
\begin{align}
g(\omega)=\begin{cases}\frac{1}{\sqrt{2\Delta\omega}}, & \omega_s-\frac{\Delta\omega}{2}\leq\omega\leq\omega_s+\frac{\Delta\omega}{2}\ \mathrm{or}\ \omega_i-\frac{\Delta\omega}{2}\leq\omega\leq\omega_i+\frac{\Delta\omega}{2};\\
0, & \mathrm{otherwise},
\end{cases}\notag
\end{align}
where $\omega_s+\omega_i=\omega_p$.

4, We use the approximation of dispersion relation with Taylor expansion around $\omega_s$ and $\omega_i$ to the first order:
\begin{align}
k_X(\omega_s+\nu)&=k_{Xs}+\frac{\nu}{v_{gXs}}, \notag\\
k_X(\omega_i+\nu)&=k_{Xi}+\frac{\nu}{v_{gXi}},\ \mathrm{where}\ X = H\ \mathrm{or}\ V,\notag
\end{align}
where $\nu\ll\omega_c$ as the bandwidth of frequency-bins is narrow. We may rewrite the Eqn.(\ref{CVState}) into:
\begin{align}
\begin{split}\ket{\psi'_{PBS}}\propto\int_{-\Delta\omega/2}^{\Delta\omega/2}\mathrm{d}\nu 
[&a^{\dag}_{H,3}(\omega_s+\nu)a^{\dag}_{H,4}(\omega_i-\nu)e^{i(\varphi_A+\nu\tau_A)}+a^{\dag}_{H,3}(\omega_i-\nu)a^{\dag}_{H,4}(\omega_s+\nu)e^{i(\varphi_B+\nu\tau_B)}\\
&a^{\dag}_{V,3}(\omega_s+\nu)a^{\dag}_{V,4}(\omega_i-\nu)e^{i(\varphi_C+\nu\tau_C)}+a^{\dag}_{V,3}(\omega_i-\nu)a^{\dag}_{V,4}(\omega_s+\nu)e^{i(\varphi_D+\nu\tau_D)}\ket{0}
\end{split}\tag{S9}\label{symstate}
\end{align}
where we put all frequency independent arguments of phases into $\varphi_k$, and denote the group delays as $\tau_k$ ($k=A,B,C,D$): \begin{align}
\begin{split}
\varphi_A&=k_pL_2+k_{Vs}L_1+k_{Hi}L_1+k_{Vs}L_3+k_{Hs}L_3'+k_{Vi}L_4+k_{Hi}L_4'\\
\varphi_B&=k_pL_2+k_{Hs}L_1+k_{Vi}L_1+k_{Vi}L_3+k_{Hi}L_3'+k_{Vs}L_4+k_{Hs}L_4'\\
\varphi_C&=k_pL_1+k_{Vs}L_2+k_{Hi}L_2+k_{Hs}L_3+k_{Vs}L_3'+k_{Hi}L_4+k_{Vi}L_4'\\
\varphi_D&=k_pL_1+k_{Hs}L_2+k_{Vi}L_2+k_{Hi}L_3+k_{Vi}L_3'+k_{Hs}L_4+k_{Vs}L_4'\\
\tau_A&=\frac{L_1}{v_{gVs}}-\frac{L_1}{v_{gHi}}+\frac{L_3}{v_{gVs}}+\frac{L_3'}{v_{gHs}}-\frac{L_4}{v_{gVi}}-\frac{L_4'}{v_{gHi}}\\
\tau_B&=\frac{L_1}{v_{gHs}}-\frac{L_1}{v_{gVi}}-\frac{L_3}{v_{gVi}}-\frac{L_3'}{v_{gHi}}+\frac{L_4}{v_{gVs}}+\frac{L_4'}{v_{gHs}}\\
\tau_C&=\frac{L_2}{v_{gVs}}-\frac{L_2}{v_{gHi}}+\frac{L_3}{v_{gHs}}+\frac{L_3'}{v_{gVs}}-\frac{L_4}{v_{gHi}}-\frac{L_4'}{v_{gVi}}\\
\tau_D&=\frac{L_2}{v_{gHs}}-\frac{L_2}{v_{gVi}}-\frac{L_3}{v_{gHi}}-\frac{L_3'}{v_{gVi}}+\frac{L_4}{v_{gHs}}+\frac{L_4'}{v_{gVs}}
\end{split}\notag
\end{align}
We substitute Eqn.(\ref{detectorfield}) and (\ref{symstate}) into (\ref{tfunction}) and find
\begin{align}
\begin{split}
\Psi(t_3,t_4)\propto\int_{-\Delta\omega/2}^{\Delta\omega/2}\mathrm{d}\nu\bigg\{&\hat{e}_3\centerdot\hat{e}_H\hat{e}_4\centerdot\hat{e}_H\{e^{i\{\varphi_A+\omega[\tau_A-(t_3-t_4)]\}}+e^{i\{\varphi_B+\omega[\tau_B-(t_3-t_4)]\}}\}\\
&\hat{e}_3\centerdot\hat{e}_V\hat{e}_4\centerdot\hat{e}_V\{e^{i\{\varphi_C+\omega[\tau_C-(t_3-t_4)]\}}+e^{i\{\varphi_D+\omega[\tau_D-(t_3-t_4)]\}}\}\bigg\}
\end{split}\tag{S10}
\end{align}

In coincidence detection we only measure wavepacket of $t_3-t_4$. As can be seen from Eqn.(S10), the biphotons are associated with four different wavepackets of coincidences. When the wavepackets perfectly overlap in space time, $\tau_A=\tau_B=\tau_C=\tau_D$ must be satisfied. The conditions are equivalent to:
\begin{align}
\tau_A+\tau_B-\tau_C-\tau_D&=(L_1-L_2+L_3-L_3'-L_4+L_4')(\frac{1}{v_{gVs}}+\frac{1}{v_{gHs}}-\frac{1}{v_{gVi}}-\frac{1}{v_{gHi}})=0\tag{S11}\label{condition1}\\
\tau_A-\tau_B+\tau_C-\tau_D&=(L_1+L_2)(\frac{1}{v_{gVs}}-\frac{1}{v_{gHs}}+\frac{1}{v_{gVi}}-\frac{1}{v_{gHi}})+(L_3+L_3'-L_4-L_4')(\frac{1}{v_{gVs}}+\frac{1}{v_{gHs}}+\frac{1}{v_{gVi}}+\frac{1}{v_{gHi}})=0\tag{S12}\label{condition2}\\
\tau_A-\tau_B-\tau_C+\tau_D&=(L_1-L_2+L_3-L_3'-L_4+L_4')(\frac{1}{v_{gVs}}+\frac{1}{v_{gVi}}-\frac{1}{v_{gHs}}-\frac{1}{v_{gHi}})=0\tag{S13}\label{condition3}
\end{align}
Eqn.(\ref{condition1}) and (\ref{condition3}) leads to $\Delta L=|L_1-L_2+L_3-L_3'-L_4+L_4'|=0$, which agrees well with our previous analysis on narrow bandwidth assumption in Eqn.(\ref{concurrencerho}). Conditions Eqn.(\ref{condition1}) and (\ref{condition3}) is equivalent to $\tau_A=\tau_C$ and $\tau_B=\tau_D$. It means that with $\Delta L=0$ the polarization entanglement will not be decoherent when $\Delta L=0$, regardless of frequency or temporal degree of freedom. 

However, one can find that, the state $\ket{\omega_s}_3\ket{\omega_i}_4$ is associated with the wavepacket centering at $\tau_A=\tau_C$, while $\ket{\omega_i}_3\ket{\omega_s}_4$ is associated with $\tau_B=\tau_D$. The leaking timing information will lead to decoherence in frequency, thus condition Eqn.(\ref{condition2}) needs to be satisfied to erase the distinguishing information. By using $\frac{1}{v_{gX}}=\left.\frac{\mathrm{d}k_X}{\mathrm{d}\omega}\right|_{\omega=\omega_{c}}\approx\frac{1}{2}(\frac{1}{v_{gXs}}+\frac{1}{v_{gXi}})$,  We may approximate the Eqn.(\ref{condition2}) to:
\begin{align}
(L_1+L_2)(\frac{1}{v_{gV}}-\frac{1}{v_{gH}})+(L_3+L_3'-L_4-L_4')(\frac{1}{v_{gV}}+\frac{1}{v_{gH}})=0\tag{S14}\label{stringentcondi}
\end{align}
This condition is shown in main text as Eqn.(10).

In practice, to avoid decoherence it requires that the time difference between wavepackets shall be much smaller than the biphoton correlation time. For example, if we assume a tranform-limited photon pair with bandwidth $\sim\Delta\omega$ of 2$\pi$$\times$100GHz ($\sim0.8$nm at telecom band), corresponding to $\sim3$ps correlation time. As can be seen from Eqn.(\ref{condition1})-(\ref{condition3}) the delay accuracy is determined by group velocity mismatch $M=\frac{1}{v_{gV}}-\frac{1}{v_{gH}}$. Assuming we can make sure that the PMF is $\delta L\sim$1cm off from the ideal case, using Eqn.(\ref{GVM}) we find the temporal walk-off error will be $\sim13$ fs, which is much smaller than the correlation time.

\end{widetext}

\end{document}